\documentclass[journal]{IEEEtran}
\ifCLASSINFOpdf
\else
\fi

\usepackage{amsmath}    
\usepackage{amsfonts}  
\usepackage{amssymb}
\usepackage{graphicx}   

\usepackage{dcolumn}
\usepackage{bm}

\newcommand{\be}{\begin{equation}}
\newcommand{\ee}{\end{equation}}
\newcommand{\bea}{\begin{eqnarray}}
\newcommand{\eea}{\end{eqnarray}}

\newcommand{\bbm}{\begin{bmatrix}}
\newcommand{\ebm}{\end{bmatrix}}

\newcommand{\rme}{{\rm{e}}}

\newcommand{\iGAj}{{j}}

\newcommand{\cliffconj}[1] { \bar{#1} }

\begin{document}
%
\title{The vector algebra war: \\ a historical perspective}
%
%
%

\author{James M.~Chappell, Azhar~Iqbal, John~G.~Hartnett, and~Derek~Abbott,~\IEEEmembership{Fellow,~IEEE}
\thanks{James M.~Chappell, Azhar~Iqbal and Derek Abbott are with the School of Electrical and Electronic Engineering, University of Adelaide, SA 5005, Australia e-mail: james.chappell@adelaide.edu.au.}
\thanks{John~G.~Hartnett is with the Institute for Photonics \& Advanced Sensing (IPAS), and the School of Physical Sciences, University of Adelaide, SA 5005 Australia.}
\thanks{Manuscript received November 13, 2015; revised: XX YY, 2015.}}

\maketitle

\begin{abstract}
There are a wide variety of different vector formalisms currently utilized in engineering and physics.  For example, Gibbs' three-vectors, Minkowski four-vectors, complex spinors in quantum mechanics, quaternions used to describe rigid body rotations and vectors defined in Clifford geometric algebra. 
With such a range of vector formalisms in use, it thus appears that there is as yet no general agreement on a vector formalism suitable for science as a whole. This is surprising, in that, one of the primary goals of nineteenth century science was to suitably describe vectors in three-dimensional space.  This situation has also had the unfortunate consequence of fragmenting knowledge across many disciplines, and requiring a significant amount of time and effort in learning the various formalisms.
We thus historically review the development of our various vector systems and conclude that Clifford's multivectors best fulfills the goal of describing vectorial quantities in three dimensions and providing a unified vector system for science.
\end{abstract}

\begin{IEEEkeywords}
Vectors, Gibbs, Hamilton, Clifford, Multivectors
\end{IEEEkeywords}

%
\IEEEpeerreviewmaketitle

\section{Introduction}
%
%
%
%
\IEEEPARstart{G}{enerally} speaking, the concept of a vector has been an extremely useful one with nearly all branches of physical science now described in the language of vectors~\cite{Crowe1967}.
Despite its great value as a concept there is nevertheless a plethora of different vector formalisms currently in use. Listed in the approximate order of their creation are:  complex numbers (planar vectors), quaternionic vectors, Gibbs' vectors, Minkowski four-vectors, complex spinors, Dirac matrix four-vectors and finally vectors defined using Clifford algebra, as well as several other vector-type formalisms.

This fact is surprising as one of the main goals of nineteenth century science was to find a suitable vector system for three-dimensional Euclidean space.  This objective was initially led by Hamilton who produced the quaternionic vectors.  Unfortunately Hamilton's system failed to live up to the initial high expectations and following an intense debate over several years, it was replaced by the Gibbs vector system in mainstream use today. We firstly identify why the quaternions fail to produce a suitable description of Cartesian vectors but also importantly identify what their natural role is.  We then show the serious failings of the Gibbs vector system before demonstrating a reconciliation of these two rival systems within Clifford geometric algebra $ C\ell(\Re^3) $.
We then conclude that this system indeed provides the most natural vector system for three-dimensional space.

\section{Analysis}
\label{analysis}

The concept of vectorial quantities actually appears to be quite an ancient one, with the parallelogram law for the addition of vectors well known to Aristotelian science from the fourth century B.C.E.~\cite{Drachmann1963} and later repeated in Newton's {\it{Principia}}.
Descartes, in 1637 however, proposed a much more radical view of vectors as quantities such that {\it `Just as arithmetic consists of only four or five operations, namely, addition, subtraction, multiplication, division and the extraction of roots....so in geometry, to find required lines it is merely necessary to add or subtract lines.'}~\cite{StuartMill1979}
This revolutionary idea was indeed successfully formulated algebraically in the nineteenth century by Wessel, Argand and Gauss.  It was achieved through the use of complex numbers $ z = a + i b $, where $ a, b \in \Re $ are real numbers and $ i = \sqrt{-1} $ is the unit imaginary. The number $ z $ was then interpreted as representing a point in the plane located at the coordinates $ [a,b] $.  This point could also be viewed as representing a vector extending out from the origin to this point. 
The ability to use a single number `$z$' to describe a two-dimensional planar point means that we now can undertake geometrical analysis in a coordinate free manner.  It is also compatible with intuition in having a single number `$z$' to refer to a single point, rather then needing to always refer to two separate coordinates. 
For example, if we have two vectors represented by the complex numbers $ z_1 $ and $ z_2 $ then vectorial addition is simply $ z_1 + z_2 $. That is, if $ z_1 = a_1 + i b_1 $ and $ z_2 = a_2 + i b_2 $ then $ z_1 + z_2 = (a_1 + a_2 ) + i (b_1 + b_2 ) $ and so satisfies the parallelogram law for adding vectorial quantities, as required. We thus have extended real numbers to a more general type of number with the addition of the imaginary component with all fundamental arithmetic operations essentially unchanged.  
Indeed, complex numbers are a division algebra and so satisfy Descartes vision of vectorial quantities being amenable to the four common arithmetic operations.  This principle of Descartes was also consistent with a later principle by Hankel for extending mathematical concepts, of the {\it the principle of the permanence of the rules of calculation}~\cite{Semadeni1984}.  

One defect of this approach to representing a Cartesian vector by a complex number is that we are setting up a real and an imaginary axis for the plane that is clearly not isotropic\footnote{Isotropy implies that for an isolated physical system experimental outcomes are independent of its orientation in space.} and so somewhat inconsistent with the principles of relativity.
Complex numbers actually describe the algebra of rotations for the plane, which is two dimensional.  Hence this has the same dimension as a two-dimensional vector space.  This is why complex numbers can doubly serve as rotation operators as well as vectors for the plane.
That is, a rotation of a vector given by a complex number $ z $ can be written
\be \label{ComplexNumberRotation}
z' = \rme^{i \theta} z .
\ee
In this case both the rotation operator $ \rme^{i \theta} $ and the vector $ z $ are represented by complex numbers.

However, following the generally successful use of complex numbers in describing vectors in the plane, researchers of the nineteenth century then turned their attention to the generalization of complex numbers to three-dimensional space in order to naturally describe these vectors.

Hamilton led this program, and in 1843 he succeeded in generalizing the complex number algebra to the quaternion algebra~\cite{Hamilton1847}. 
A quaternion can be written
\be \label{quaternionFull}
q = a + v_1 \boldsymbol{i} + v_2 \boldsymbol{j} + v_3 \boldsymbol{k} ,
\ee
where the three basis vectors have a negative square $ \boldsymbol{i}^2 = \boldsymbol{j}^2 = \boldsymbol{k}^2 =  -1 $ and are anticommuting with each other. 
Hamilton's quaternions form a four-dimensional associative division algebra over the real numbers represented by $ {\mathbb{H}}  $.  Once again being a division algebra, like complex numbers, they are amenable to all the common arithmetic operations.  The sequence of algebras $ \Re $, $ \mathbb{C} $ and $ \mathbb{H} $ are constructed to be division algebras, that is, they are closed with an inverse operation.  Indeed, the required algebraic rules for quaternions follow from this closure property~\cite{Doran2003}.
These properties also, in fact, make them naturally suited to describe rotations in space as they also have the closure property.

Quaternions are also isotropic, as required, with a three dimensional `vector' represented as $ \boldsymbol{v} = v_1 \boldsymbol{i} + v_2 \boldsymbol{j} + v_3 \boldsymbol{k} $.
Hamilton then claimed that being a generalization of complex numbers to three dimensions it would therefore logically be the appropriate algebra to describe three-dimensional space.  Indeed he proposed, many years before Einstein or Minkowski, that if the scalar $a$, in Eq.~(\ref{quaternionFull}), was identified with time then the four-dimensional quaternion can be a representation for a unified spacetime framework~\cite{Graves1882}.  Indeed, when Minkowski came to develop the idea of a unified spacetime continuum, after considering the merits of quaternions, he chose rather to extend the Gibbs vector system with the addition of a time coordinate to create a four-component vector~\cite{Minkowski1908}.

Using quaternions, we have rotations in three dimensions given by the bilinear form 
\be 
q' = \rme^{ \boldsymbol{a}/2 } q \rme^{  -\boldsymbol{a}/2 } ,
\ee
where $ \boldsymbol{a} = a_1 \boldsymbol{i} + a_2 \boldsymbol{j} + a_3 \boldsymbol{k} $ is a vector quaternion describing the axis of rotation.
One of the objections raised against the quaternions at the time was their lack of commutativity, however in order to describe rotations in three dimensions this is actually a requirement for the algebra as three dimensional rotations themselves are non-commuting in general.  Indeed this form of three-dimensional rotation is extremely useful as it avoids problems such as gimbal lock and has other advantages such as being very efficient in interpolating rotations.

It thus appeared at first that quaternions may indeed be the ideal algebra for three-dimensional physical space that had been sought. Unfortunately there was one cloud on the horizon, the fact that a vector quaternion squares to the negative Pythagorean length.
Indeed, Maxwell commented on this unusual fact, noting that the kinetic energy, which involves the square of the velocity vector, would therefore be negative~\cite{Crowe1985}. 
Maxwell, despite these reservations, formulated the equations of electromagnetism in quaternionic form.  Maxwell, however, backed away from a complete endorsement of the quaternions, recommending in his treatise on electricity and magnetism 
 {\it `the introduction of the ideas, as distinguished from the operations and methods of Quaternions'}~\cite{Maxwell1873}.

The difficulties with quaternions led to a breakaway formalism of the Gibbs vector system. Gibbs considered that the most useful function of the quaternions was in their forming of the dot product and the cross product operations.  That is, for two vector quaternions $ \boldsymbol{v} =  v_1 \boldsymbol{i} + v_2 \boldsymbol{j} + v_3 \boldsymbol{k} $  and $ \boldsymbol{w} =  w_1 \boldsymbol{i} + w_2 \boldsymbol{j} + w_3 \boldsymbol{k} $ we find through expanding the brackets, that
\bea \label{quaternionExpansion}
\boldsymbol{v} \boldsymbol{w} & = & (v_1 \boldsymbol{i} + v_2 \boldsymbol{j} + v_3 \boldsymbol{k})(w_1 \boldsymbol{i} + w_2 \boldsymbol{j} + w_3 \boldsymbol{k}) \\ \nonumber
& = & -v_1 w_1 - v_2 w_2 - v_3 w_3 + (v_2 w_3 - v3 w_2 )\boldsymbol{i} \\ \nonumber
& & + (v_3 w_1 - v1 w_3 )\boldsymbol{j}+ (v_1 w_2 - v2 w_1 )\boldsymbol{k} \\ \nonumber 
& = & - \boldsymbol{v} \cdot \boldsymbol{w} +  \boldsymbol{v} \times \boldsymbol{w} , \nonumber
\eea
using the negative square and the anticommuting properties of the basis vectors.  We can also see how the vector quaternion squared $ \boldsymbol{v}^2 = - \boldsymbol{v} \cdot \boldsymbol{v} $ is thus the negative of the Pythagorean length as noted previously.
We can see though how indeed the dot and cross products naturally arise from the product of two vector quaternions.  
Gibbs then considered that adopting the separate operations of the dot and cross products acting on three-vectors could thus form the basis for a more efficient and straightforward vector system.

This led to an intense and lengthy debate over several years between the followers of Gibbs and the followers of Hamilton, beginning in 1890, over the most efficient vectorial system to be adopted in mathematical physics~\cite{Crowe1967}. The supporters of Hamilton were able to claim that quaternions being generalized complex numbers were clearly preferable as they had a proper mathematical foundation.  The Gibbs' side of the debate though argued that the non-commutativity of the quaternions added many difficulties to the algebra compared with the much simpler three-vector formalism in which the dot and cross products were each transparently displayed separately.
Ultimately, with the success of the Gibbs formalism in describing electromagnetic theory, exemplified by the developments of Heaviside~\cite{Heaviside1925}, and with an apparently more straightforward formalism, the Gibbs system was adopted as the standard vector formalism to be used in engineering and physics. 
This outcome to the debate is perhaps surprising in hindsight as in comparison with quaternions, the Gibbs vectors do not have a division operation and two new multiplication operations are required beyond elementary algebra and so therefore did not satisfy the basic principles of Descartes or Hankel.
The trend of adopting the Gibbs vector system, however, continued in 1908 when Minkowski rejected the quaternions as his description of spacetime and chose to rather extend the Gibbs three-vector system, on the grounds that the quaternions were too restrictive.  In fact the Lorentz boosts are indeed difficult to describe with quaternions and interestingly this difficulty arises from the square of vector quaternions being negative---the same problem that was identified earlier by Maxwell.

Also in 1927 with the development of quantum mechanics and the Schr{\"o}dinger  and Pauli equations, a vector in the form of the complex spinor was adopted to describe the wave function, despite the fact that the spinor is in fact isomorphic to the quaternion!  However with the arrival of the Dirac equation in 1928, which required an eight-dimensional wave function, the four-dimensional quaternions were then clearly deficient. The quaternions, though, can be complexified that results in an eight-dimensional algebra. However, in 1945, Dirac concluded that the true value of quaternions lies in their unique algebraic properties, and so generalizing the algebra to the complex field is therefore not the best approach~\cite{Dirac1944}.

In summary, it therefore appears that Hamilton had indeed been wrong with his assertion that his quaternions were the natural algebra for three-dimensional space, as they were ultimately deficient in describing the Dirac equation and had difficulties with Lorentz transformations.
As we have already noted, complex numbers define the algebra of rotations for the plane and indeed Hamilton had only generalized this algebra to produce the algebra of rotations in three dimensions.  This fact demonstrates that Gibbs, in fact, was correct in asserting that quaternions were not suitable to describe Cartesian vectors in three-dimensions due to their natural role as rotation operators.  Indeed, the more natural identification of the quaternions as the even subalgebra of $ C\ell(\Re^3 ) $, and hence rotation operators, is illustrated in Table~\ref{tableGroups}.

The solution to this dilemma would therefore appear to be to add Hamilton’s quaternion algebra to the Gibbs Cartesian vectors in some way in order to provide a complete algebraic description of space---a goal that was indeed achieved by Clifford.
The Clifford geometric algebra $ C\ell(\Re^3 ) $, being an eight-dimensional linear space, is indeed able to subsume the quaternion algebra and the Gibbs vectors into a single formalism, as required.  We now therefore describe the Clifford geometric algebra in three dimensions.

\section{Clifford's vector system}
\label{clifford}

A Clifford geometric algebra $ C\ell \left(\Re^n\right) $ defines an associative real algebra over $ n $ dimensions~\cite{Clifford1878}. In three dimensions the algebra $ C\ell \left(\Re^3\right) $ forms an eight-dimensional linear space and is isomorphic to the $ 2 \times 2 $ complex matrices. 
For three dimensions, we adopt the three quantities $ e_1, e_2, e_3 $ for basis vectors\footnote{While it is possible to define Clifford geometric algebra in a coordinate free manner, it is more illustrative to follow Hamilton's approach and firstly define a basis.} that are defined to anticommute, so that $ e_1 e_2 = - e_2 e_1 $, $ e_1 e_3 = - e_3 e_1 $, and $ e_2 e_3 = - e_3 e_2 $, but unlike the quaternions these quantities square to positive one\footnote{The basis elements of a Clifford algebra can be defined more generally to have a positive or a negative square and over any number of dimensions.  For example, a Clifford algebra is typically defined over several hundred dimensions in order to analyze signals in terahertz spectroscopy.~\cite{Xie2011}}, that is $ e_1^2 = e_2^2 = e_3^2 = 1 $. We thus have a vector 
\be
\boldsymbol{v} = v_1 e_1 + v_2 e_2 +v_3 e_3 
\ee
that now has a positive square $ \boldsymbol{v}^2 = v_1^2 + v_2^2 + v_3^2 $, giving the Pythagorean length.  This thus avoids Hamilton's defect with the vector quaternions having a negative square.

Using the elementary algebraic rules just defined, multiplying two vectors produces
\be \label{CliffordProductRaw}
\boldsymbol{v} \boldsymbol{w} = \boldsymbol{v} \cdot \boldsymbol{w} + \iGAj \boldsymbol{v} \times \boldsymbol{w} ,
\ee
where $ \iGAj = e_1 e_2 e_3 $.  Thus Clifford's vector product forms an invertible product, as well as producing the dot and cross products in the form of a complex-like number.
Indeed, this allows us to write the dot and cross products as
\bea
\boldsymbol{v} \cdot \boldsymbol{w} & = & \frac{1}{2} \left ( \boldsymbol{v} \boldsymbol{w} + \boldsymbol{w} \boldsymbol{v} \right ), \\ \nonumber \iGAj \boldsymbol{v} \times \boldsymbol{w} & = & \frac{1}{2} \left ( \boldsymbol{v} \boldsymbol{w} - \boldsymbol{w} \boldsymbol{v} \right ) \nonumber
\eea
so that they become simply the symmetric and antisymmetric components of a more general Clifford product.  This also becomes a convenient way to the define the wedge product as $  \boldsymbol{v} \wedge \boldsymbol{w} =  \nonumber \iGAj \boldsymbol{v} \times \boldsymbol{w} $.
For the inverse of a vector we therefore have  
\be
\boldsymbol{v}^{-1} = \boldsymbol{v}/\boldsymbol{v}^2 . 
\ee
That is, $ \boldsymbol{v} \boldsymbol{v}^{-1} = \boldsymbol{v} \boldsymbol{v}/\boldsymbol{v}^2 = \boldsymbol{v}^2/\boldsymbol{v}^2 = 1 $, as required.
Rotations use a similarly efficient form to the quaternions with
\be \label{CliffordRotations}
\boldsymbol{v}' = \rme^{-\iGAj \boldsymbol{a}/2 } \boldsymbol{v} \rme^{\iGAj \boldsymbol{a}/2 } ,
\ee
where $  \boldsymbol{a} $ is the axis of rotation.
Also similarly to quaternions we can combine scalars and the various algebraic components into a single number called a multivector
\be \label{multivectorFull}
 a + v_1 e_1 + v_2 e_2 + v_3 e_3 + w_1 e_2 e_3 + w_2 e_3 e_1 + w_3 e_1 e_2 + b e_1 e_2 e_3 .
\ee
Now using $ j = e_1 e_2 e_3 $ we can form the dual relations $ j e_1 = e_2 e_3 $, $ j e_2 = e_3 e_1 $ and $ j e_3 = e_1 e_2 $.  Therefore we can write the multivector as
\be \label{multivectorWithVectors}
M = a + \boldsymbol{v} + j \boldsymbol{w} + j b ,
\ee
with the vectors $ \boldsymbol{v} = v_1 e_1 + v_2 e_2 + v_3 e_3 $ and $ \boldsymbol{w} = w_1 e_1 + w_2 e_2 + w_3 e_3 $.
We also define Clifford conjugation on a multivector $ M = a + \boldsymbol{v} + j \boldsymbol{w} + j b $ as
\be
\cliffconj{M} = a - \boldsymbol{v} - j \boldsymbol{w} + j b 
\ee
that gives a general multivector inverse $ M^{-1} = \cliffconj{M}/M \cliffconj{M} $.

In order to restore isotropy to complex numbers and the Argand plane, we can introduce the Clifford geometric algebra $ C\ell(\Re^2 ) $ where the complex numbers are isomorphic to the even subalgebra. The multivector in $ C\ell(\Re^2 ) $ describes separately the rotation algebra and vectors. That is, we have a multivector
\be
a + v_1 e_1 + v_2 e_2 + i b ,
\ee
where $ e_1 , e_2 $ are now isotropic anticommuting basis vectors, where $ e_1^2 = e_2^2 = 1 $ and $ i = e_1 e_2 $, the bivector of the plane.  We can then write for the rotation of a planar vector
\be
v' = \rme^{-i \theta} v ,
\ee
where now we have the Cartesian vector $ v= v_1 e_1 + v_2 e_2  $ acted on by the even subalgebra $ r \rme^{-i \theta} $, where $ i = e_1 e_2 $.
We can see the equivalence with the complex number formula given earlier because we can write a Cartesian vector $ v = e_1 (v_1 + e_1 e_2 v_2 ) = e_1 z $ that then reverts to the complex number rotation formula in Eq.~(\ref{ComplexNumberRotation}).
Hence $ C\ell(\Re^2 ) $ is a more precise way to describe the Cartesian plane.  The identification of the complex numbers as the even subalgebra of $ C\ell(\Re^2 ) $ is illustrated in Table~\ref{tableGroups}.

Clifford's system is also mathematically quite similar to Hamilton's system, with the main distinction being that the basis vectors have a positive square.  Unlike Hamilton's system, we can also form the compound quantities with the basis vectors such as bivectors $ e_1 e_2, e_3 e_1, e_2 e_3 $ and trivectors $ j = e_1 e_2 e_3 $. This greater dimensionality allows the inclusion of the Gibbs Cartesian vectors and the quaternions as a subalgebra. Indeed it can be shown that quaternions are isomorphic to the even subalgebra of the multivector, with the mapping $ {\boldsymbol{i}} \leftrightarrow e_2 e_3 $, $ {\boldsymbol{j}} \leftrightarrow e_1 e_3 $, $ {\boldsymbol{k}} \leftrightarrow e_1 e_2 $ and the Gibbs vector can be replaced by the vector component of the multivector as shown in Eq.~(\ref{multivectorFull}).  The fact that the quaternions form the even subalgebra and so are represented by the bivectors explains why the vector quaternions have the  property of squaring to the negative of the Pythagorean length.  The bivectors are in fact pseudovectors or axial vectors and so clearly not suitable for their use as polar vectors as Hamilton claimed.

While Hamilton's non-commutivity was one of the things that counted against his vector system at the time it actually is exactly what is needed in a vector system in three dimensions, as three-dimensional rotations are intrinsically non-commuting. 
A further oversight of the Gibbs vector system is that while attempting to describe linear vectorial quantities his system completely ignored the presence of directed areal and volume objects within three-dimensional space.  This situation is also remedied with the Clifford multivector that algebraically describes the points, lines, areas and volumes of three-dimensional space.  The four geometric elements also describe the four types of physical variables of scalar, polar vector, axial vectors and pseudoscalars.
 So in fact both Gibbs and Hamilton fell well short of the ultimate objective of algebraically describing three-dimensional space and was only completed by Clifford.  

Thus, the generalization of quaternions to $ C\ell(\Re^3 ) $, which includes the quaternions as the even subalgebra, provides a more complete generalization of the two-dimensional plane than quaternions and a more efficient algebraic description of three-dimensional space.

\begin{figure}
\includegraphics[scale=0.47]{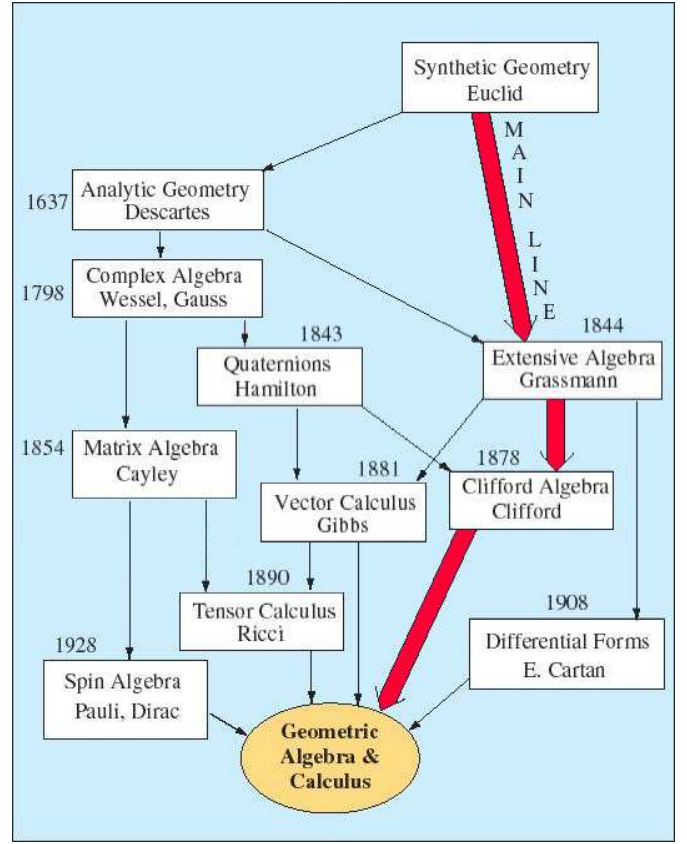}
\caption{\label{DescentOfVectors} The descent of the various vector systems. The main path of development beginning from Euclid geometry down through Grassman and then to Clifford. Other parallel developments using complex numbers, quaternions, Gibbs' vectors, tensors, matrices and spinor algebra subsumed into the general formalism of Clifford geometric algebra with the inclusion of calculus.}
\end{figure}

Note that, while we have typified the three key vector systems with the name of their main originator, many other scientists were involved in their development.  For the Gibb's vector system one of the key developers and proponents was Oliver Heaviside who applied it very effectively to problems in electrical engineering. For Hamilton's system of quaternions, they were further developed and championed by Peter Tait.  Clifford~\cite{Clifford1878} acknowledged the key theoretical work of the algebra of extension developed previously by Hermann Grassmann, with Clifford's system further developed and popularized more recently by David Hestenes~\cite{Hestenes:1966,GA,GA2}. The path of development of the various vectorial systems that culminate in Clifford geometric algebra is shown in Fig~\ref{DescentOfVectors}

\subsection{Comparison of physical theories}
\label{comparison}

Within Clifford's multivectors we write a spacetime event as 
\be
S = t + x_1 e_1 + x_2 e_2 + x_3 e_3.
\ee
We then find
\bea \nonumber
S \cliffconj{S} & = & (t + x_1 e_1 + x_2 e_2 + x_3 e_3)(t - x_1 e_1 - x_2 e_2 - x_3 e_3) \\ 
& = & t^2 - x_1^2 - x_2^2 - x_3^2 ,  
\eea
thus producing the required metric.  This is equivalent to Minkowski's four-vectors~\cite{Minkowski1908} $ s = [t,x_1,x_2,x_3 ] $ and where $  s \cdot \cliffconj{s} = t^2 - x_1^2 - x_2^2 - x_3^2  $.

Lorentz boosts also follow immediately from the formalism simply by the exponentiation of vectors similar to the exponentiation of bivectors for rotations, as shown in Eq.~(\ref{CliffordRotations}), that is
\be \label{CliffordBoost}
S' = \rme^{\boldsymbol{v}/2} S \rme^{\boldsymbol{v}/2} ,
\ee
where $ \boldsymbol{v} $ describes the boost direction. The form of the expression in Eq.~(\ref{CliffordBoost}) is not available with quaternions as we require vectors with a positive square and is also not available with Gibbs' vectors because they do not allow us to form an exponential series.
In order to achieve this with four-vectors we need to generate a $ 4 \times 4 $ boost matrix to act on the four-vector.
With Clifford's system we can also neatly summarize the special Lorentz group with the operator $ \rme^{\boldsymbol{v} + j\boldsymbol{w}} $ that in a single expression describes rotations and boosts.  Alternate descriptions of spacetime have been developed, using Clifford algebra, that are isomorphic to this formulation such as the algebra of physical space (APS)~\cite{baylis2004geometry,baylis2004relativity}, or the space-time algebra (STA)~\cite{Hestenes:1966}.

Using the Gibbs vector system we can reduce Maxwell's field equations down to the following four equations
\bea \label{MaxwellClassical}
\mathbf{\nabla} \cdot \mathbf{E} & = & \frac{\rho}{\epsilon},   \,\,\,\,\,\,\, \rm{(Gauss\text{'} \,\, law)} ; \\ \nonumber
\mathbf{\nabla} \times \mathbf{B} - \frac{1}{c^2} \frac{\partial \textbf{E} }{\partial t} & = & \mu_0 \mathbf{J},  \,\,\,  \rm{(Amp\grave{e}re\text{'}s \,\, law)} ; \\ \nonumber
\mathbf{\nabla} \times \mathbf{E} +  \frac{\partial \mathbf{B}}{\partial t}   & = &  \mathbf{0} , \,\,\,\,\,\,\,\,\, \rm{(Faraday\text{'}s \,\, law)} ; \\ \nonumber
\mathbf{\nabla} \cdot \mathbf{B}  & = &  0 , \,\,\,  \,\,\,  \,\,\,  \rm{(Gauss\text{'} \,\, law \,\, of \,\, magnetism)} ,  \nonumber
\eea
where $ \nabla = e_1 \partial_x + e_2 \partial_y + e_3 \partial_z $.  This form of the electromagnetic equations is found today in most modern textbooks~\cite{Griffiths:1999}.

Using Clifford geometric algebra, Maxwell's four equations can be written with the single equation~\cite{Chappell2014IEEE,Baylis2001}
\be
\left (\partial_t + \nabla \right) F = \frac{\rho}{\epsilon} - \mu c \mathbf{J} ,
\ee
where the field is $  F = \boldsymbol{E} + \iGAj c \boldsymbol{B}  $ and $ \nabla = e_1 \partial_x + e_2 \partial_y + e_3 \partial_z $.
Thus the Clifford vector system allows a single equation over the reals as opposed to four equations required using the Gibbs vector system.  This simplified form for Maxwell's equations follows from the unification of the dot and cross products into a single algebraic product~\cite{Friedman2002}.  However beyond this lack of economy of representation other issues arise within the Gibbs formalism.

The Gibbs vector system defines both the electric field $ \mathbf{E} = [E_1, E_2, E_3] $ and the magnetic field $ \textbf{B} = [B_1, B_2, B_3] $ as three component vectors.  However, the magnetic field has different transformational properties to the electric field, which are normally taken into account through referring to the electric field as a {\itshape{polar}} vector and the magnetic field as an {\itshape{axial}} vector. As stated by Jackson: 
{\it We see here....a dangerous aspect of our usual notation. The writing of a vector as `$\mathbf{a}$' does not tell us whether it is a polar or an axial vector}~\cite{Jackson1998}.
The Clifford vector system correctly distinguishes the electric and magnetic field as a vector and bivector respectively within the single field variable $ F = \boldsymbol{E} + \iGAj c \boldsymbol{B} $, thus being an improvement over the Gibbs notation.

Using a mixture of notation, the spinor, four-vector, complex numbers and matrix notation we can write the Dirac equation
\be
\gamma^{\mu} \partial_{\mu} \psi = -i m  \psi ,
\ee
where the Dirac gamma matrices satisfy $ \gamma^\mu \gamma^\nu + \gamma^\nu \gamma^\mu = 2 \eta^{\mu \nu} I $, where $ \eta^{\mu \nu} $ describes the Minkowski metric signature $ (+,-,-,-) $.  The four gradient being $ \gamma^0 \partial_0 + \gamma^1 \partial_1 + \gamma^2 \partial_2 + \gamma^3 \partial_3 $, where $ \partial_0 \equiv \partial_t \equiv \frac{\partial}{\partial t} $.  The wave function is a four component complex vector $ \psi = [ z_1, z_2 , z_3 , z_4 ] $, where $ z_{\mu} $ are complex numbers.  As we can see a whole suite of new notation needs to be introduced in order to describe the Dirac equation.

Now, the Dirac wave-function is in fact eight-dimensional and so naturally corresponds with the eight-dimensional Clifford multivector.  Indeed, we can write the Dirac equation in Clifford's formalism~\cite{Lounesto2001} as
\be
\partial M = - \iGAj m M^* e_3,
\ee
where the trivector $ j = e_1 e_2 e_3 $ replaces the unit imaginary allowing us to stay within a real field.  The four-gradient is defined as $ \partial = \partial_t + \nabla $ and the involution $ M^* = a - \boldsymbol{v} + j \boldsymbol{w} - j b $. 
Thus the Clifford multivector, without any additional formalism such as matrices or complex numbers, naturally describes the central equations of electromagnetism and quantum mechanics over a real field.  

As already noted, Gibbs developed his system by splitting the single quaternion product into the separate dot and cross products.
This is reflected in the standard calculus results for differentiation
\bea
\frac{d}{dt} \left ( \boldsymbol{a} \cdot  \boldsymbol{b} \right ) & = & \frac{d \boldsymbol{a} }{dt} \cdot  \boldsymbol{b} + \boldsymbol{a}  \cdot \frac{d \boldsymbol{b} }{dt} , \\ \nonumber 
\frac{d}{dt} \left ( \boldsymbol{a} \times  \boldsymbol{b} \right ) & = & \frac{d \boldsymbol{a} }{dt} \times \boldsymbol{b} + \boldsymbol{a}  \times \frac{d \boldsymbol{b} }{dt} . \nonumber
\eea
Clifford (and Hamilton) naturally unifies these results into the single expression
\be
\frac{d}{dt} \left ( \boldsymbol{a} \boldsymbol{b} \right ) = \frac{d \boldsymbol{a} }{dt} \boldsymbol{b} + \boldsymbol{a}  \frac{d \boldsymbol{b} }{dt} .
\ee
Also the div and curl relations are unified  into a single expression 
\be
\nabla \boldsymbol{a} = \nabla \cdot \boldsymbol{a} + \iGAj \nabla \times \boldsymbol{a} .
\ee
Clifford geometric algebra and the geometric product allow more generalizations in the field of calculus, such as unifying the Gauss and Stokes theorem~\cite{GA}.

\subsection{Bound vectors---Pl{\"{u}}cker coordinates}
\label{plucker}

Vectors are often typified as quantities with a magnitude and a direction and commonly represented as arrows acting from the origin of a coordinate system.  However, we would like to generalize this idea to include vectors that are not acting through the origin.  This will then allow us to naturally represent a quantity such as torque as a force offset from the origin, for example.  This generalization can be achieved as an extension of Gibbs' vector system through defining a  Pl{\"{u}}cker coordinate, which extends a normal three-vector of Gibbs to a six-dimensional vector, where the three extra components represent the vector offset from the origin~\cite{Dorst2009}.  

Within Clifford's system this idea can be represented more easily through the use of the vector and bivector components of the multivector
\be
V = \boldsymbol{v} + j \boldsymbol{w} = \boldsymbol{v} + \boldsymbol{v} \wedge \boldsymbol{r},
\ee
where $ \boldsymbol{v} $ is the direction of the normal free vector and $ \boldsymbol{r} $ is the offset of this vector from the origin.  Thus if the wedge product $ \boldsymbol{v} \wedge \boldsymbol{r} = j \boldsymbol{v} \times \boldsymbol{r} = 0 $ then this implies that the vectors are parallel and so implies that the vector passes through the origin and so we are reverting to a pure position vector.  Hence this naturally generalizes the normal concept of a position vector.

For example if we have two force vectors $ F_1 = \boldsymbol{f}_1 +  j \boldsymbol{t}_1 =  \boldsymbol{f}_1 + \boldsymbol{f}_1 \wedge \boldsymbol{r}_1 $ and $ F_2 = \boldsymbol{f}_2 +  j \boldsymbol{t}_2 =  \boldsymbol{f}_2 + \boldsymbol{f}_2 \wedge \boldsymbol{r}_2 $, where each Clifford six-vector consists of two forces $ \boldsymbol{f}_1 $ and $ \boldsymbol{f}_2 $ which are each offset from the origin by $ \boldsymbol{r}_1 $ and $ \boldsymbol{r}_2 $ respectively.  We then find the trivector part of the Clifford product $ F_1 F_2 $ is
\be
 j (\boldsymbol{f}_1 \cdot \boldsymbol{t}_2 + \boldsymbol{f}_2 \cdot \boldsymbol{t}_1) .
\ee
Now, we would expect, the trivector part will give the torsion of these two forces, which is indeed the case, with the sign of the product indicating a left or right hand screw direction.  If this product is zero then the two forces are coplanar and there is no net twist force.  Hence the Clifford multivector allows a natural extension to bound vectors not available with either quaternions or Gibbs vectors without significant additions to the notation.

\begin{table*}[ht]
	\centering
\renewcommand{\arraystretch}{1.3}
\caption{The algebra of space for two and three dimensions given $ C\ell(\Re^2) $ and $ C\ell(\Re^3) $ respectively.  These contain Cartesian vectors and their rotational algebra that is given by the even subalgebra. }  \label{tableGroups}

\begin{tabular}{|l|l|l|l|}
\hline
Dimension  & Clifford & Even subalgebra (Rotations) & Vector  \\
\hline \hline
 2 & $ C\ell(\Re^2) $ & Complex numbers  & $ \boldsymbol{v} = v_1 e_1 + v_2 e_2  $  \\
 &  &  $  a + i b \in {\mathbb{Z}} $ &  $ i = e_1 e_2 $ \\ \hline
3 & $ C\ell(\Re^3) $  & Quaternions &  $ \boldsymbol{v} = v_1 e_1 + v_2 e_2 + v_3 e_3 $ \\
 &   & $ q_0 + q_1 \boldsymbol{i} + q_2 \boldsymbol{j} + q_3 \boldsymbol{k} \in {\mathbb{H}} $  &  $ j = e_1 e_2 e_3 $ \\
\hline 
\end{tabular}
\end{table*}

\section{Discussion}
\label{discussion}

The first main defect of the Gibbs vector system is the lack of an inverse operation due to the splitting of the Clifford product into the separate dot and cross products. 
Also describing a plane using the orthogonal vector is only workable in three dimensions. That is, an orthogonal vector does not exist in two dimensions and in four dimensions and higher there is an infinitude of orthogonal vectors for a given plane. It is actually more natural to define an areal quantity as lying in the plane of the two vectors under consideration, as is the case with Clifford's system.  This then allows planar quantities to be represented uniformly in an arbitrary number of dimensions.

For example, the Gibbs system can find the area of a parallelogram from the magnitude of the cross product $  \boldsymbol{a} \times \boldsymbol{b} $ where the direction of the vector formed by the cross product is orthogonal to the plane formed by the vectors $ \boldsymbol{a} $ and $ \boldsymbol{b} $. The volume of a parallelepiped is typically found from the scalar triple product $  \boldsymbol{a} \cdot ( \boldsymbol{b} \times  \boldsymbol{b}) $.

The Clifford product of two vectors $ \boldsymbol{a} $ and $ \boldsymbol{b} $, on the other hand is $ \boldsymbol{a} \boldsymbol{b} $
where the magnitude of the bivector component gives the area formed by the two vectors.  In this case, bivectors are a native descriptor of area that give the orientation of the plane it represents.  
The trivector components of the Clifford product $ \boldsymbol{a} \boldsymbol{b} \boldsymbol{c} $ will give the volume of the parallelepiped and is an obvious extension of the area formula as volume intuitively relates to the trivector components.  Thus while the Gibbs formulas are essentially unmotivated, the Clifford product intuitively produces areal and volume quantities.

In Clifford geometric algebra, in order to ascertain the geometric relationship between two vectors $ \boldsymbol{a} $ and $ \boldsymbol{b} $ we can simply form the product
\be
 \boldsymbol{a} \boldsymbol{b} =  \boldsymbol{a} \cdot \boldsymbol{b} +  \boldsymbol{a} \wedge \boldsymbol{b} .
\ee
If the scalar component $ \boldsymbol{a} \cdot \boldsymbol{b}  = 0 $ then the vectors are orthogonal, or if the bivector or area is zero, then they are parallel.  The Clifford vector product thus provides a simple but general way to compare the orientation of two vectors in a single expression.
If seeking to determine whether or not three vectors in space lie in a plane we can simply check the trivector component of  $ \boldsymbol{a} \boldsymbol{b} \boldsymbol{c} $ and if it is zero then they are coplanar.   

More generally, because Clifford vectors, such as $ \boldsymbol{v} $ are now elementary algebraic quantities all common functions are available such as logarithms, trigonometric, exponential functions as well as the general calculations of roots.  For example, the $ \sin \boldsymbol{v} $ or $ \log \boldsymbol{v} $ or even the expressions such as $ 2^{\boldsymbol{v}} $ of raising a number to a vector power, can now be calculated~\cite{Chappell2015}.

It is interesting that electrical engineers, such as Heaviside, led the adoption of the original Gibbs' vector system. Is it possible that forward thinking electrical engineers will also lead the adoption of Clifford's vector system?

\section{Conclusion}

We conclude that the mathematical formalism developed by Clifford of $ C\ell(\Re^3) $ is the natural algebra to describe three-dimensional space.  Clifford geometric algebra $ C\ell(\Re^3) $  thus fulfills this important goal of nineteenth century science.   Clifford's geometric algebra also extends seamlessly to spaces with an arbitrary number of dimensions $ C\ell(\Re^n) $, where $ n $ is an integer, and so is scalable. That is, the algebra in two dimensions $ C\ell(\Re^2) $ naturally extends to $ C\ell(\Re^3) $ and up to $ C\ell(\Re^n) $ in an orderly manner.  Each additional dimension doubling the size of the space.

In attempting to find the natural algebraic description of physical space Hamilton generalized the complex number algebra to the four-dimensional quaternion algebra. We noted though that the quaternions only provide the algebra describing rotations in three dimensions and so do not provide suitable Cartesian vectors and so is not a complete description.  The confusion between the rotational algebra and Cartesian vectors can actually only arise in three dimensions in which we have both three translational degrees of freedom and three rotational degrees of freedom.  This simple fact was thus at the basis of the misunderstanding between Hamilton and Gibbs. The resolution of the confusion is supplied by the Clifford geometric algebra $ C\ell(\Re^3) $ that absorbs both the quaternions and complex numbers as subalgebras as well as including a Cartesian vector component, as shown in Eq.~(\ref{multivectorWithVectors}) and in Table~\ref{tableGroups}.

We also showed how Clifford's system provides a very natural formalism for Maxwell's electromagnetism, special relativity and the Dirac equation.  The competing formalism of quaternions requires the addition of complex numbers and the Gibb's vector system requires the addition of matrices, spinors and complex numbers.  
As the Clifford trivector $ j = e_1 e_2 e_2 $ is a commuting quantity having a negative square, it can replace the abstract unit imaginary and allows all relationships to be described over a strictly real space~\cite{BaylisWhyI1992,GA2,Doran2003}.
We also showed how Clifford multivectors can describe both free and bound vectors efficiently thus providing a generalization of vectors not found in either competing system.

Thus the goal of the nineteenth century to find the natural algebraic description of three-dimensional space was achieved by Clifford with eight-dimensional multivectors, as shown in Eq.~(\ref{multivectorWithVectors}) and as we have shown can supersede to a large degree the various vector systems used today.  
New applications for Clifford algebra are steadily appearing in many fields of engineering and physics such as electromagnetism~\cite{Baylis:1992}, optics~\cite{Sugon2004}, Fourier transforms~\cite{Hitzer2013}, terahertz spectroscopy~\cite{Xie2011}, satellite navigation~\cite{ablamowicz2004lectures}, robotics~\cite{Hildenbrand2008}, computer graphics and computer vision~\cite{Dorst2009,Sommer2001,Berthier2014}, quantum mechanics~\cite{Hestenes,CIL}, quantum computing~\cite{chappellGrover2012}, special and general relativity~\cite{baylis2004,Doran1993}.



%




\ifCLASSOPTIONcaptionsoff
  \newpage
\fi

\begin{IEEEbiography}{James M. Chappell}
BE (civil, 1984), Grad.Dip.Ed.~(1993), B.Sc.~(Hons, 2006) and PhD (quantum computing, 2011) from the University of Adelaide, began his career in civil engineering followed by employment as a computer programmer before retraining as a school teacher.  He then returned to academia, completing a science degree, followed by a PhD in quantum computing.  During his PhD he specialized in quantum computing and Clifford's geometric algebra. Dr~Chappell is presently a Visiting Scholar at the School of Electrical \& Electronic Engineering, at the University of Adelaide, working on applications of geometric algebra.  
\end{IEEEbiography}

\begin{IEEEbiography}{Azhar Iqbal}
received the degree in physics from the University of Sheffield, U.K., in 1995
and received the Ph.D. degree in applied mathematics from the University of Hull, U.K., in 2006. In 2006, he won the Postdoctoral Research Fellowship for Foreign Researchers from the Japan Society for the Promotion of Science (JSPS) to work under Prof. T. Cheon at the Kochi University of Technology, Japan.  In 2007, he won the prestigious Australian Postdoctoral (APD) Fellowship from the Australian Research Council, under Derek Abbott, at the School of Electrical and Electronic Engineering, the University of Adelaide. For two years took up a post as an assistant professor at the Department of Mathematics and Statistics, King Fahd University of Petroleum and Minerals, Dhahran, Saudi Arabia. Dr Iqbal is currently an adjunct senior lecturer with the School of Electrical and Electronic Engineering, The University of Adelaide.
\end{IEEEbiography}

\begin{IEEEbiography}{John G.~Hartnett} received a B.Sc.~(Hons) and Ph.D.~(with distinction)
in Physics from The University of Western Australia (UWA), Crawley,
W.A., Australia. He served as a Research Professor with the Frequency
Standards and Metrology Research Group, at the The University of Western
Australia. In 2013, he took up an Australian Research Council (ARC)
Discovery Outstanding Researcher Award (DORA) fellowship based within
the Institute for Photonics \& Advanced Sensing, the School of Physical Sciences, 
the University of Adelaide, where he is currently an
Associate Professor. His research interests include planar
metamaterials, development of ultra-stable microwave oscillators based
on sapphire resonators, and tests of fundamental theories of physics
such as special and general relativity using precision oscillators.
Prof.~Hartnett was the recipient of the 2010 IEEE UFFC Society
W.~G.~Cady Award. He was a corecipient of the 1999 Best Paper Award
presented by the Institute of Physics Measurement Science and
Technology.
\end{IEEEbiography}


\begin{IEEEbiography}{Derek Abbott} was born in South Kensington, London, UK, in 1960.  He received a 
B.Sc.~(Hons) in physics from Loughborough University, U.K. in 1982 and the Ph.D. degree in electrical and electronic engineering from the University of Adelaide, Adelaide, Australia, in 1995, under K.~Eshraghian and B.~R.~Davis. From 1978 to 1986, he was a research engineer at the GEC Hirst Research Centre, London, U.K. From 1986--1987, he was a VLSI design engineer  at Austek Microsystems, Australia. Since 1987, he has been with the University of Adelaide, where he is presently a full Professor with the School of Electrical and Electronic Engineering.  He holds over 800 publications/patents and has been an invited speaker at over 100 institutions. Prof.~Abbott is a Fellow of the Institute of Physics (IOP) and a Fellow of the IEEE from 2005. He has won a number of awards including the South Australian Tall Poppy Award for Science (2004), the Premier's SA Great Award in Science and Technology for outstanding contributions to South Australia (2004), an Australian Research Council (ARC) Future Fellowship (2012), and the David Dewhurst Medal (2015). He has served as an Editor and/or Guest Editor for a number of journals including {\sc IEEE~Journal~of~Solid-State~Circuits}, {\it Journal~of~Optics~B} (IOP),  {\it Microelectronics~Journal} (Elsevier), {\it PLOSONE}, {\sc Proceedings of the IEEE}, and the {IEEE Photonics Journal}.  He is currently on the editorial boards of {\sc IEEE Access}, {\it Scientific Reports} (Nature), and {\it Royal Society Open Science} (RSOS). Prof~Abbott co-edited {\it Quantum~Aspects~of~Life,} Imperial College Press (ICP), co-authored {\it Stochastic~Resonance,} Cambridge University Press (CUP), and co-authored {\it Terahertz Imaging for Biomedical Applications,} Springer-Verlag.

Prof.~Abbott's interests are in the areas of multidisciplinary physics and electronic engineering applied to complex systems. His research programs span a number of areas of stochastics, game theory, photonics, biomedical engineering, and computational neuroscience.
\end{IEEEbiography}




\end{document}